\documentclass[aps,prb,twocolumn,showpacs]{revtex4}
\usepackage{tabularx,graphicx}
\usepackage{epsfig}

\newcommand{\tFM}{t_{\rm Fe-Mo}}
\newcommand{\tMo}{t_{\rm Mo-Mo}}
\newcommand{\tg}{{\bf t}_{\rm 2g}}
\newcommand{\eg}{{\bf e}_{\rm g}}
\newcommand{\LMFO}{{\rm Sr}_{ 2-x}{\rm La}_{ x}{\rm Fe Mo
O}_6}
\newcommand{\LWFO}{{\rm Sr}_{ 2-x}{\rm La}_{ x}{\rm Fe W
O}_6}
\newcommand{\LWFOo}{{\rm Sr}_{\rm 2}{\rm Fe W
O}_6}
\newcommand{\eMo}{\epsilon_{\rm Mo}}
\newcommand{\eFe}{\epsilon_{\rm Fe}}

\newcommand{\LCMO}{{\rm La}_{ 1-x}{\rm Ca}_{ x}{\rm Mn O}_3}
\newcommand{\ex}{e^{i {\bf k}_x a}}
\newcommand{\ey}{e^{i {\bf k}_y a}}
\newcommand{\cex}{e^{- i {\bf k}_x a}}
\newcommand{\cey}{e^{- i {\bf k}_y a}}
\newcommand{\ek}{\epsilon_{\bf k}}

\begin{document}

\bibliographystyle{apsrev}
\author{A. Taraphder$^{1,2}$, and F. Guinea$^1$}
\title{Orbital ordering and magnetic structures in $\LMFO$ and $\LWFO$ double perovskites.}
\affiliation{$^1$ Instituto de Ciencia de Materiales de Madrid,
CSIC, Cantoblanco, E-28043 Madrid, Spain \\$^2$ Department of
Physics and Center for Theoretical Studies, Indian Institute of
Technology, Kharagpur 721302, India}
\begin{abstract}
We analyzed the possible magnetic and orbital orderings of double
perovskites, using a simple extension of the double exchange model
well suited for these compounds. Orbital ordering is favored by
the on site repulsion at the Fe ions. We obtain a rich phase
diagram, including ferri- and antiferromagnetic phases, which can,
in turn, be  metallic or insulating, depending on the existence of
orbital order.
\end{abstract}

\maketitle

\section{Introduction}
Ferromagnetic double perovskites of composition $\LMFO$ (where La
can be substituted by other double valency ions), known since some
time\cite{SW72} have attracted a great deal of attention recently,
because of their high Curie temperature, metallic character, large
magnetoresistance and potential
applications\cite{Ketal98,Ketal99,Setal00,Getal00,Retal00,Betal01,
Tetal02,Vetal02}. While lattice distortions and phonons do not
seem to play a major role, disorder and in particular the antisite
defects play a significant role in the properties of these
compunds\cite{Oetal99,Setal00b,Getal01,Setal01,Metal01,Getal01b,Setal02b,FF04}.
Effects of doping have also been extensively studied\cite{Metal00,
Netal01,Setal02,Fetal03,Fetal03b,Wetal04,Wetal04b,Netal04}
recently . Related family of compounds based on $\LWFOo$ are, on
the other hand, insulating and
antiferromagnetic\cite{Ketal99b,FTK01}. Solid solutions including
Mo and W ions have also been
studied\cite{Letal01,Letal02,Detal02}.

The magnetic and electronic structures of the double perovskites
with composition $\LMFO$ admit a simple description\cite{Setal00}.
The Mo$^{5+}$ and Fe$^{3+}$ ions are located at the alternate
nodes of a simple cubic lattice. The strong exchange coupling
within the d orbitals of the Fe ion leads to the formation of a
spin 5/2 moment, from electrons which occupy the exchange-split
two $\eg$  and the three $\tg$ orbitals. The three remaining three
$\tg$ orbitals at the Fe sites are hybridized, through the
intermediate O ion, with the $\tg$ orbitals of the neighboring Mo
ions. Because of the symmetry of these orbitals, hopping between
them can only take place along two of the three lattice axes,
leading to three decoupled two dimensional bands. Band structure
calculations suggest that the direct hopping between the Mo $\tg$
orbitals is not negligible (this hopping does not change the two
dimensional nature of the bands). The number of electrons in the
conduction band, per unit cell, is ${\rm x} = 1+x$ where $x$ is
the concentration of the divalent La$^{2+}$ ions. Most of the
calculations reported below are given in terms of the band
filling, ${\rm x}$, and the correspondence with the doping level,
$x$, is highlighted when needed.

These arguments suggest that the minimal description of the
electronic structure of these materials requires three parameters:
the energy difference between the $\tg$ levels in the Fe and Mo
ions, $\Delta = \eMo - \eFe$, the Fe-Mo hopping, $\tFM$, and the
Mo-Mo hopping, $\tMo$. We assume that the exchange splitting
between the spin $\tg$ states in the Fe ions is much larger than
the width of the conduction band. Hence, we need to consider one
spin state at each Fe ion. This truncation of the states used to
describe the conduction band is similar to the one used in
reducing the two band ferromagnetic Kondo lattice to the single
band double exchange model, used in the study of $\LCMO$ and
related compounds. This tight binding model can also be used to
study the $\LWFOo$ compound\cite{FTK01}.

The phase diagram of the hybridized Mo-Fe system described above
has been studied using the dynamical mean field
method\cite{PCM03}, and within an exact treatment of the
electronic wavefunction which allows for the study of
defects\cite{Aetal03}. Using realistic values for the parameters
the Curie temperature is comparable to the one experimentally
observed. The generic features of the phase diagram have some
resemblance to those of double exchange model, to which it can be
reduced in some limits. One of the most remarkable differences is
the appearance of an antiferromagnetic phase even in the absence
of direct antiferromagnetic interactions\cite{Aetal03}.

The analysis outlined above implicitly assumes the equivalence of
the three $\tg$ orbitals at the Fe and Mo sites, and it neglects
the possibility of non trivial types of orbital order. Electron
electron interactions may break the symmetry between the $\tg$
orbitals. Such phases have been extensively investigated in the
related double exchange compounds of composition $\LCMO$ and
related materials\cite{BKK99,HFD01,MT02,MT03,MT04,Aetal03b,B04}.
The electron electron interaction in $\LMFO$ can be large, as the
conduction band has a significant weight on the Fe orbitals. In
the present paper, we analyze the possibility of the breaking of
the orbital symmetry in the $\LMFO$ family of compounds, using
mean field theory. The electron-electron interactions are
described by means of a local repulsion term, $U$, between
electrons in different orbitals at the same Fe site. Orbital
ordering has already been reported for the stoichiometric $\LWFOo$
compound\cite{FTK01}. We obtain a rich phase diagram, as a
function of the number of electrons in the conduction band and the
difference between the Fe and Mo levels. The method is discussed
in the next section. Then, we present the main results. Section IV
discusses the relevance of our findings for the physical
properties of these materials.
\section{The model and the method of calculation.}
\begin{figure}[!]
  \begin{center}
       \epsfig{file=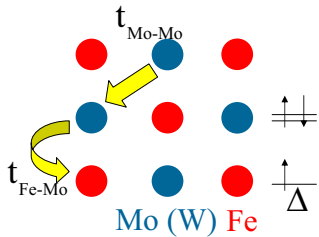,height=7cm}
    \caption{(Color online). Sketch of the interactions included
in ${\cal H}_{\rm kin}$, eq.(\protect{\ref{Hkin}}).}
    \label{hopping}
\end{center}
\end{figure}
We assume that the conduction band arises from the hybridization
of $\tg$ orbitals at the Mo and Fe sites. There are three spin
polarized orbitals at the Fe ion, and six orbitals at the Mo ion.
The hopping takes place within planes determined by the symmetry
of the orbitals (for instance, the hybridization does not allow
for hopping between ${\rm d}_{\rm xy}$ orbitals along the $z$
axis). There is hopping only between states with the same spin.
The levels at the Fe sites are shifted with respect to those at
the Mo sites by an energy $\Delta$. Within these approximations,
the kinetic energy of the conduction electrons is described by
three independent hamiltonians of the form:
\begin{widetext}
\begin{equation}
{\cal H}_{\rm kin}^{\alpha=1,2,3} = \tFM \sum_{\rm i,j n.n. , s}
c_{\alpha,i,s}^\dag d_{\alpha,j,s} + \tMo \sum_{\rm ij n.n.n. , s}
c_{\alpha,i,s}^\dag c_{\alpha,j,s} + \Delta_\alpha \sum_{\rm i,s}
d_{\alpha,i,s}^\dag d_{\alpha,i,s} \label{Hkin} \end{equation}
\end{widetext}
The operators $c_{\alpha,i,s}$ create electrons with spin $s$ in
the orbital $\alpha$ at a Mo site, $i$, and the operators
$d_{\alpha,i,s}^\dag$ have the same effect at Fe sites. Note that
the index $s$ is either $1$ or $-1$ at the Fe sites, while it can
take both values at the Mo sites. We have conserved the spin index
at the Fe sites in order to describe antiferromagnetic phases
along (111) planes, like that observed in $\LWFOo$ (see below). A
schematic representation of the couplings in eq.(\ref{Hkin}) is
shown in Fig.[\ref{hopping}].
\begin{figure}[!]
  \begin{center}
       \epsfig{file=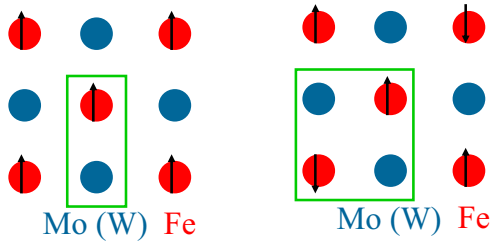,height=7cm}
    \caption{(Color online). Configurations of the core spins at the
    Fe ions considered in the text. The rectangles denote the corresponding
unit cell.}
    \label{spin_texture}
\end{center}
\end{figure}

The hamiltonian ${\cal H}_{\rm kin}$ is defined on a square
lattice. The resulting energy bands depend on the configuration of
the core spins at the Fe sites. We will consider the two textures
shown in Fig.[\ref{spin_texture}]. When all core spins are
parallel, the unit cell includes one Fe and one Mo site. The
subbands with spins antiparallel to the core spins are built up
from one orbital at the Fe site and another orbital at the Mo(W)
site. The energies, $\ek$, are given by:
\begin{widetext}\begin{equation} {\rm Det} \left|
\begin{array}{cc} 4 \tMo \cos ( {\bf k}_ x a ) \cos ( {\bf k}_y a
) - \ek &\tFM \ex \ey \\ \tFM \cex \cey &\Delta_\alpha - \ek
\end{array} \right| = 0\end{equation} \end{widetext} The band with
spin parallel to the Fe core spins is localized at the Mo sites.
The energy is:
\begin{equation}
\ek =  4 \tMo \cos ( {\bf k}_ x a ) \cos ( {\bf k}_y a )
\end{equation}
In the antiferromagnetic configuration shown in the right side of
Fig.[\ref{spin_texture}] the unit cell includes four sites. The
spin up and spin down bands are degenerate, although they are
localized in different regions of the lattice. The spin up band is
derived from two orbitals at the Mo sites, and the orbital at the
Fe site whose core has spin down. The energies are given by:
\begin{widetext}\begin{equation} {\rm Det} \left|
\begin{array}{ccc} - \ek &\tMo ( 1 + \ex ) ( 1 + \ey ) &\tFM ( 1 +
\cex ) \\ \tMo ( 1 + \cex ) ( 1 + \cey ) &- \ek &\tFM ( 1 + \cey )
\\ \tFM ( 1 + \ex ) &\tFM ( 1 + \ey ) &\Delta_\alpha - \ek \end{array}
\right| = 0
\end{equation} \end{widetext}
We also include a repulsion between electrons in different
orbitals at the Fe sites. This term has the form:
\begin{equation}
{\cal H}_{\rm int} = U \sum_{\rm i, \alpha \ne \alpha'}
d_{i,\alpha,s}^\dag d_{i,\alpha,s} d_{i,\alpha',s}^\dag
d_{i,\alpha',s}
\end{equation}
We use the Hartree-Fock method to analyze the effect of this term.
Then, assuming that the occupancies of the different orbitals at
the Fe site are not the same, the Fe levels are given by:
\begin{equation}
\Delta_\alpha = \eFe - \eMo + U \sum_{\alpha' \ne \alpha} \langle
d_{i,\alpha',s}^\dag d_{i,\alpha',s} \rangle
\label{levels}\end{equation} Here, $\eFe$ and $\eMo$ are defined
in the absence of interaction corrections. The values of the
expectation values $\langle d_{i,\alpha',s}^\dag d_{i,\alpha',s}
\rangle$ have to be calculated selfconsistently, inserting the
levels given in eq.(\ref{levels}) into eq.(\ref{Hkin}).

Note that the filling of the conduction bands, ${\rm x}$ can vary
between 0 and 9, $0 \le {\rm x} \le 9$. In the following we will
show calculations for band fillings within this range, although
fillings such that ${\rm x} \ge 2$ cannot be obtained by
substituting the divalent Sr ions by trivalent ions.

From the total energy versus doping curves (see below) we identify
regions of phase separation in the phase diagram using the Maxwell
construction. This construction allows us to identify densities at
which different phases can coexist as they have the same chemical
potential.
\section{Results.}
In the following, we use as unit of energy $\tFM$. From band
structure calculations\cite{Setal00}, this parameter is $\tFM
\approx 0.3$eV. We also fix $\tMo = 0.2 \tFM$ and $U = 12 \tFM$.
These values are less well defined, but consistent with existing
band structure calculations\cite{Setal00}. The results do not
change qualitatively for other values of these parameters,
although the finer details are modified. We vary $\Delta = \eFe -
\eMo + U/3$ from 0 (strongly hybridized band, probably adequate
for $\LMFO$) to $\Delta = - 4 \tFM$ (conduction electrons mostly
localized on Fe ions, valid for $\LWFO$). The filling of the
conduction band can vary from 0 to 9. Experimentally relevant
values of band filling (${\rm x} $) lie between 1 and 2,
corresponding to doping levels, $x$, such that $0 \leq x \leq 1$.
\begin{figure}[!]
  \begin{center}
       \epsfig{file=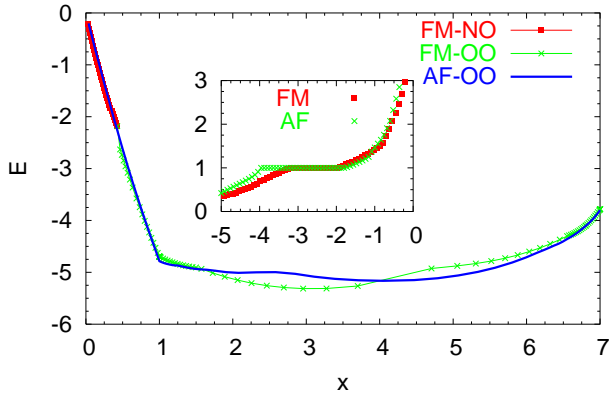,height=6cm}
    \caption{(Color online). Total energy (per site)
    of the ferrimagnetic solutions with and without
    orbital ordering and for the antiferromagnetic solution (orbitally
    orderered) as a
    function of band filling, ${\rm x}$. The splitting between the Fe and Mo levels
    is $\Delta = \eFe - \eMo + U/3 = -4 \tFM$.
    Inset: band filling (vertical axis) as function of chemical
    potential (horizontal axis) in units of $\tFM$. Note that the
    doping is equal to the band filling minus one. OO stands for orbitally ordered,
    and NO for orbital degeneracy. FM and AF stand for the ferrimagnetic and
    antiferrromagnetic phases discussed in the text.}
    \label{D-4}
\end{center}
\end{figure}

Results for the total energy of selfconsistent solutions for
$\Delta = -4 \tFM$ are shown in Fig.[\ref{D-4}]. This case favors
the localization of the conduction electrons at the Fe sites, a
situation which probably describes well $\LWFO$\cite{FTK01}. The
energy difference (per site) of the different phases is much
smaller than the electronic kinetic energy. The most stable phase
around $x=0$ is the antiferromagnetic phase with orbital ordering.
One of the three $\tg$ orbitals at the Fe site is occupied, while
the two others are pushed to higher energies. This phase is
insulating, as shown in the kink of the energy vs. $x$ curve, and
the finite range of values of the chemical potential compatible
with the band filling equal to one (doping $x=0$), shown in the
inset. This range gives the gap in the electronic spectrum, which
is of order $1-2 \times \tFM$. Our calculations suggest a wide
range of phase coexistence, between the stoichiometric case, with
doping $x=0$, and  doping $x \approx 1$. At sufficiently high
dopings, an orbitally ordered ferromagnetic phase is stable.

The nature of the insulating phase in this regime is in agreement
with the band structure calculations in\cite{FTK01}. One of the
bands with spin antiparallel to the core Fe spin lies separated by
a gap from the others. This band can accommodate one electron, and
doping beyond $x=0$ requires the filling of states above this gap.
\begin{figure}[!]
  \begin{center}
       \epsfig{file=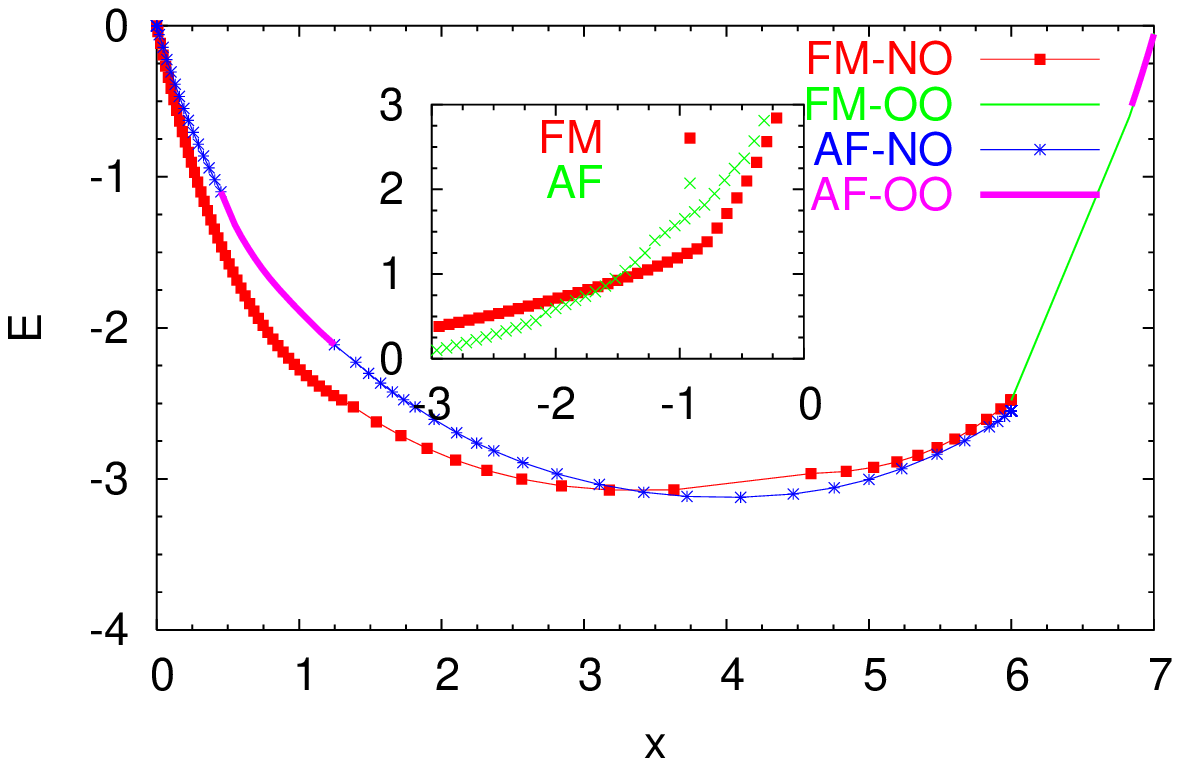,height=6cm}
    \caption{(Color online). As in figure[\protect{\ref{D-4}}], but for
    $\Delta = 0$.}
    \label{D0}
\end{center}
\end{figure}

The opposite case, where the Fe and Mo orbitals are optimally
hybridized ($\Delta = 0$), is shown in Fig.[\ref{D0}]. This choice
of parameters is probably adequate for $\LMFO$. The most stable
phase near $x=0$ is a metallic ferrimagnetic phase, without
orbital ordering. At sufficiently high dopings, however, the
antiferromagnetic, orbitally ordered phase prevails. These results
are consistent with the calculations in\cite{Aetal03}, although
the antiferromagnetic phase is shifted towards larger values of
$x$, due to the different value of $\tMo$ used here. There is a
first order phase transition between these two phases, with a
significant range of doping values for which phase separation
takes place. The two competing phases are metallic throughout the
physically relevant doping range.
\begin{figure}[!]
  \begin{center}
       \epsfig{file=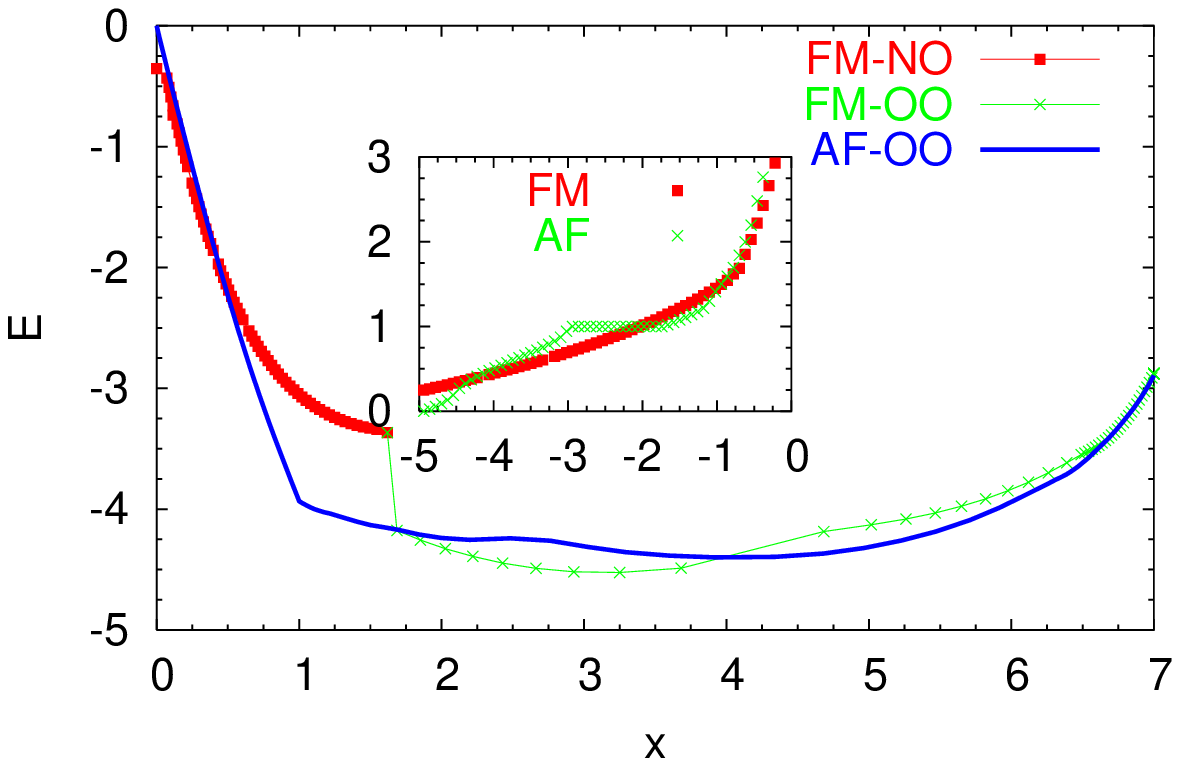,height=6cm}
    \caption{(Color online). As in figure[\protect{\ref{D-4}}], but for
    $\Delta = -3 \tFM$.}
    \label{D-3}
\end{center}
\end{figure}
\begin{figure}[!]
  \begin{center}
       \epsfig{file=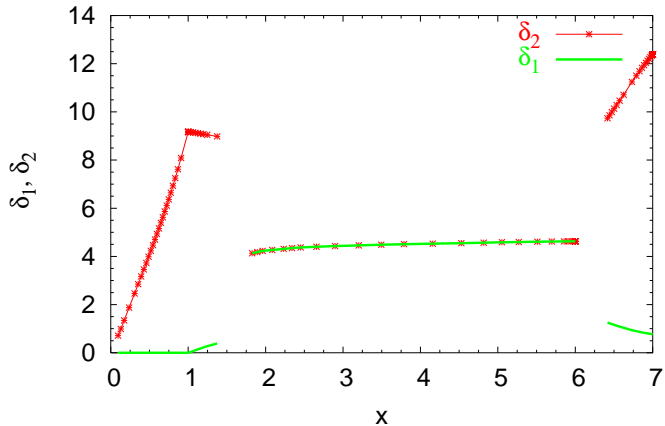,height=6.5cm}
    \caption{(Color online). Splitting of the orbital levels
    at the Fe ions for $\Delta = -2 \tFM$
     as function of band filling, ${\rm x}$, in units of $\tFM$.}
    \label{OP}
\end{center}
\end{figure}
The most complex intermediate situation between these two regimes
takes place for $\Delta = -3 \tFM$, as shown in Fig.[\ref{D-3}].
The orbitally ordered  antiferromagnetic phase competes with the
ferrimagnetic phase with and without orbital order. At low
dopings, $x \leq 0$, the ferrimagnetic phase is the most stable.
The insulating, orbitally ordered antiferromagnetic phase prevails
near zero doping, $x=0$, and the ferrimagnetic phase with orbital
order has the lowest energy for $x \ge 0.6$. There are sizable
regions of phase separation around $x=0$.

The difference between the orbital levels at the Fe ion, which can
be used to characterize the orbital order is shown in
Fig.[\ref{OP}]. At low band fillings, when orbital order exists,
the level splitting is, roughly, $\delta_1 - \delta_2 \approx U
x$. This effect leads, for $x=1$ to a sizeable gap between the
highest occupied and the lowest empty bands, as discussed above.
\begin{figure}[!]
  \begin{center}
       \epsfig{file=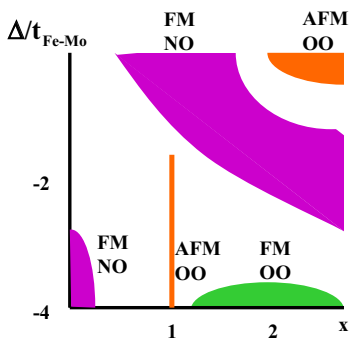,height=7cm}
    \caption{(Color online). Schematic view of the phase diagram as function of
    the separation between the Fe and Mo(W)levels and the band filling ${\rm x}$. FM and AF
    denote the ferrimagnetic and antiferromagnetic phases described in the text. OO stands
    for orbital order, and NO labels a phase without orbital order. The
    antiferromagnetic phase at ${\rm x}=1$ is insulating. Phase separation takes place
    in the blank regions.}
    \label{phased}
\end{center}
\end{figure}

The phase diagram which arises from these calculations is shown in
Fig.[\ref{phased}]. There is a competition between the different
phases considered here, with regions of phase separation.
\section{Conclusions.}
We have studied the phase diagram of a tight binding model
compatible with the known results about the band structure of
$\LMFO$ and $\LWFO$. The model used resembles the double exchange
model proposed for the manganites. The presence of a non magnetic
ion (Mo,W) with levels at similar energies to the orbital levels
of the magnetic ion (Fe) induces significant changes in the
resulting magnetic structures. The large (threefold) degeneracy of
the $\tg$ levels involved also influences the phase diagram.

The presence of non magnetic ion favors phases without net
magnetization, if the band filling is such that the conduction
electrons tend to reside in the non magnetic ion. This situation
can be relevant for $\LMFO$ at sufficiently large dopings. At
lower dopings (band filling ${\rm x} \rightarrow 0$), and in a
strongly hybridized band, the material tends to be a ferromagnetic
metal, through a mechanism similar to double exchange.

Orbital order is favored when the conduction electrons are mostly
localized in the magnetic, highly correlated ion (Fe). Then, a
single band splits from the rest, leading to a system which is
insulating at stoichiometry, $x=0$ (band filling ${\rm x} = 1$).
This phase can be either ferrimagnetic or antiferromagnetic, with
a slight tendency towards the latter. This situation may describe
$\LWFO$. We have not analyzed the suppression of orbital order at
finite temperatures. The existence of a large gap in the
insulating phase adequate for $\LWFO$ implies that, within the
Hartree Fock approximation, orbital order persists above the
N\`eel temperature. This insulating phase is separated by regions
of phase separation from other stable phases at different dopings.
A similar situation arises in the single band Hubbard model at
half filling\cite{Getal00b}.

The orbital ordering may be difficult to observe in a system which
tends to be highly disordered, such as $\LMFO$ and $\LWFO$.
Orbital order favors anisotropic electronic properties, at each of
the $\tg$ orbitals selects a plane perpendicular to one of the
axes of the cubic lattice. The formation of domains because of
antisites or other types of disorder can restore isotropy,
although the system should be anisotropic at short length scales.
\section{Acknowledgments.}
We are thankful to J. A. Alonso, L. Brey, L. A. Fern\'andez, J.
Fontcuberta, M. Garc{\'\i}a-Hern\'andez, V. Mart{\'\i}n-Mayor and
D. D. Sarma for many helpful conversations and comments during the
course of this work. We are also thankful to Ministerio de Ciencia
y Tecnolog{\'\i}a (Spain) for financial support through grant
MAT2002-0495-C02-01. One of us (A. T.) is thankful to Ministerio
de Educaci\'on y Cultura (Spain) for a sabbatical stay.
\bibliography{MoFe_U_3}
\end{document}